# An Analytic Model for Nano Confined Fluids Phase-Transition
## (Applications for Confined Fluids in Nanotube and Nanoslit)

By

**Tahmineh (Ezzat) Keshavarzi**[(1)], **Rezvan Sohrabi**[(1)] and **G.Ali Mansoori**[(2)]

(1). Isfahan University of Technology, Isfahan, Iran
(2). University of Illinois at Chicago, Chicago, IL USA

## Abstract

In this report, an analytic model to predict phase transitions of confined fluids in nano systems is presented and it is used to predict the behavior of the confined fluid in nanotubes and nanoslits. In our approach besides including a third degree of freedom due to wall effect to define the state of the system, the tensorial character for pressure is considered. Using the perturbation theory of statistical mechanics it is shown that the van der Waals equation of state is equally valid for small as well as large systems. The model proposed is shown to predict the liquid-vapor phase transition as well as the critical point in any size confined fluid systems. It is also shown that the critical temperature increases with the size of the nano system and finally it reaches the macroscopic critical temperature value as the diameter of the nanotube (or width of the nanoslit) approaches infinity. The proposed model can also demonstrate the existence of the local density and phase fragmentations during phase transitions in a confined nano system.

________________________________
Email Addresses:
T. Keshavarzi:      *keshavrz@cc.iut.ac.ir*
G.A. Mansoori:      *mansoori@uic.edu*
R. Sohrabi:         *sohrabi@aut.ac.ir*

[2]

# Introduction:

Studies into the behavior of fluids in confined small/nano systems are presently an area of research interest through experimental, theoretical and computer simulation methods[1-5]. Such systems have some present and many envisioned future applications in the emerging fields of nanoscience and nanotechnology. Therefore understanding the properties of confined fluids in nano systems, which differ significantly from the bulk fluids, is of fundamental and practical importance.

In the present report we limit ourselves to fluids (liquid and vapor) that may exist in a confined-closed cylindrical nanopore (nanotube) or in a nanoslit. The structural and dynamical properties of a fluid, which is confined in a nanopore, or nanoslit, may differ significantly from a macroscopic fluid system due to the geometry of the confinement and also due to significance of molecular interactions with the walls as compared to intermolecular interactions between the confined fluid molecules[6].

In analogy to large systems the whole fluid in a nanotube, or nanoslit, may exist in a one-phase or a two-phase state, depending on the governing independent state variables. Of course, for a confined fluid in a macroscopic system there are a maximum of two governing independent variables, such as, for example, temperature and pressure.

For a small confined-fluid system the number of independent variables is expected to be more than two due to the appreciable number of particles close to the wall compared to bulk particles and the non-uniformity of fluid density. The principles of phase separation / transition are well-defined and formulated in macroscopic systems[7-10] known as the thermodynamic limit [($N$ and $V$)→∞ but $N/V$=finite]. However, for small systems consisting of limited number of particles, principles governing the separation / transition of phases are not well understood yet. Actually the thermodynamic property relations in small systems are also not yet formulated [14].

Thermodynamic property relations in small /nano systems are functions of the geometry and internal structure of the system under consideration. In contrast to thermodynamics of large systems, the thermodynamic properties and property relations of small systems will be generally different in different environments[5]. Understanding of phase transition in systems composed of finite number of particles is a peculiar and unsolved problem from, both, the theoretical and experimental points of view. Considering that nanoscale systems consist of finite number of particles (intermediate in size between isolated atoms, or molecules, and bulk materials) principles of phase transitions need to be reformulated for nano systems.

Nano systems composed of limited number of particles and confined to nanoscale volumes are not large enough compared to the range of the interatomic and intermolecular forces existing between their particles. A more precise characterization of all such small systems is to call them non-extensive systems[5,11,12,14].

Due to the lack of proper experimental techniques we have not been able to directly study phase transitions in nano systems. However, investigations either through indirect measurement methods, or by computer simulation techniques, have been successful to distinguish various phases of matter in small systems, as small as clusters of a few atoms or molecules[6,13,14].

Nano, or small, systems are those whose linear dimension is of the characteristic range of the interaction between the particles comprising the system. Of course, astrophysical systems are also in this category and from the point of view of interactions between planets and stars they may be considered "small". Even though the principles of phase transitions are not well defined for small systems, there are many phenomena in small systems that resemble the phase transitions in large systems. This has been specially the case in the study of clusters of tens, hundreds and thousands of molecules[6,13,14].

Considering that nanoscale systems consist of finite number of particles (intermediate in size between isolated atoms / molecules and bulk materials), principles of phase transitions, as formulated for large systems, need to be reformulated for nano systems[14]. While we can have control-mass (closed) nano systems, however due to the fact that such systems are not in thermodynamic limit and they may not be in equilibrium, from the point of view of Gibbsian thermodynamics, we may not be able to use the Gibbs phase rule for such systems. Also, for nano systems, which are actually non-extensive, the definition and separation of extensive and intensive variables is not quite clear, and very possibly, they depend on the size of the system[12,14].

In what follows in this report we first prove the validity of the van der Waals equation of state for small / nano systems. Then with consideration of the tensorial character of pressure in fluids confined in nano systems and the internal energy equation we develop directional analytic equations of state for nano confined fluids. Solution of the resulting equations of state provides us with phase transition characteristics of confined nano fluids.

## Proof of validity of the van der Waals Equation of State for Small Systems:

J.D. van der Waals in 1873 worked on, probably, the first recorded predictive theory of phase transitions which resulted in the well known van der Waals equation of state (vdW eos)[15],

$$(P + \frac{aN^2}{V^2})(V - Nb) = NkT , \qquad (1)$$

Where $P$ is the otal pressure, $V$ is the total volume, $T$ is the absolute temperature, $N$ is the number of molecules, $k$ is the Boltzmann constant and $a$ & $b$ are constants. This equation, when solved for pressure versus volume, gives a simple and satisfactory account of fluid phase transitions. The vdW eos is now modified in various forms and its modifications have found many applications in the analysis of thermodynamic properties of pure fluids and mixtures in macroscopic systems of scientific and industrial interest. It's theoretically based extensions and modifications have been also the subject of many theories of statistical mechanics[16-20].

Molecular basis of the vdW eos is through the assumption of a simple intermolecular potential energy function and its application in the perturbation expansion of the Helmholtz free energy. In deriving the vdW eos the following intermolecular potential energy function is used as it is also shown graphically in Figure (1):

$$\phi(r) = \begin{cases} \infty & for \ \ r \leq \sigma \\ -\dfrac{\varepsilon}{r^n} & for \ \ r > \sigma \ \ and \ \ n \to \infty \end{cases} \qquad (2)$$

Through the application of the perturbation theory[17,21], the Helmholtz free energy of a system of $N$ particles interacting via a pair potential $\phi(r_{12})$ can be written as an expansion around the ideal gas properties by the following equation[1].

$$A = A^{ig} - (\frac{kTN^2}{2V^2}) \int \int \{ \exp[-\phi(r_{12}) / kT] - 1 \} dr_1 \, dr_2, \qquad (3)$$

where $N$ is the total number of particles and $V$ is the total volume of the system.

For a finite-volume nano system we may split the integral into two regions, $0 < r_{12} < \sigma$ and $\sigma < r_{12}$. Then the above perturbation expansion will assume the following form[1]

$$A = A^{ig} - (\frac{kTN^2}{2V^2}) \left[ \iint\limits_{0 < r_{12} < \sigma} \left[ e^{-\phi(r_{12})/kT} - 1 \right] dr_1 \, dr_2 + \iint\limits_{\sigma < r_{12}} \left[ e^{-\phi(r_{12})/kT} - 1 \right] dr_1 \, dr_2 \right] \qquad (4)$$

By inserting Eq.(2) in (4) the first integral will reduce to $-(4/3)\pi V \sigma^3$. Then since $\phi(r) = -\dfrac{\varepsilon}{r^n}$ is very small when $n \to \infty$ the intgrand in the second integral will reduce to $e^{-\phi(r_{12})/kT} - 1 = \dfrac{\varepsilon}{kTr^n}$. Then following a similar algebraic manipulation as Zarragoicoechea and Kuz[1] we can conclude the van der Waals equation for the Helmholtz free energy of a nanosystem,

$$A = Nf(T) - NkT \ln(V - Nb) + \frac{N^2 a}{V} \qquad (5)$$

This indicates that the van der Waals equation of state is valid for small, as well as large, systems. Of course, the numerical values of parameters $a$ and $b$ of the vdW eos may vary with the size of the system. From the Helmholtz free energy expression, Eq. (5), we can derive the chemical potential based on the VdW EOS,

$$\mu = \left[\frac{\partial A}{\partial N}\right]_{T,V} = f(T) - kT \ln(V - Nb) + \frac{kTNb}{V - Nb} - \frac{2aN}{V}. \qquad (6)$$

Having proven the validity of the vdW eos for small systems it can be used for thermodynamic property prediction of small systems. It is understood that the vdW eos is not quantitatively accurate due to the approximation in the intermolecular potential energy function, Eq. (2), and the perturbation expansion, Eq. (3). However, qualitatively for comparison purposes between small and large systems this eos can be quite useful.

## An analytic Model to Predict the Phase Behavior of Fluids in Nano Systems

The state of a confined (control mass) substance in a macroscopic system is defined using up to two degrees of freedom[7]. For a confined substance in a nano / small system we may need to define its state using more than two degrees of freedom[6]. This difference, between confined macroscopic and confined nano systems, is due to the relatively much larger number of molecules adjacent to walls with respect to bulk molecules of small systems as compared to a macroscopic system. As a result, in nano systems the interaction of molecules with their external environment (or walls) will have to be taken into account. Due to wall effects it is believed that additional degrees of freedom will be necessary to define the state of a confined nano system. In this article all the wall effects of a confined fluid in a uniform and constant diameter nanotube, or constant thickness nanoslit, are incorporated in one additional degree of freedom. Therefore in our approach we add an additional term resulting from the surface effects, to the Helmholtz free energy of the system. Also, in order to consider the confinement effect, which alters the average density to local density in the space, we assume the pressure is a diagonal tensor, $p$, with components $p_{ii}(i = x, y, z)$ similar to the procedure followed by Zarragoicoechea and Kuz[1]. In this way, we may obtain the equation of state (EOS) of a fluid in a nanotube or nanoslit.

Similar to the procedure introduced by Zarragoicoechea and Kuz[1] we represent the internal energy, $E$, in which pressure has tensorial character by the following equation[1,12]

$$dE = TdS - \sum_i p_{ii} d\delta_{ii} V ,$$ (7)

where $T$ is the absolute temperature, $S$ is the entropy and $V$ is the volume of the system. The second term on the right-hand side represents the work done by the internal tension under a specific deformation $d\delta_{ii}$ in the volume $V$. Then the Helmholtz free energy, will be written as[1]

$$A = N\mu - \sum_i p_{ii} d\delta_{ii} V ,$$ (8)

where $N$ is the total number of molecules and $\mu$ is the molecular chemical potential. Eq.s (7) and (8) are general and applicable for any size system. For nano-confined systems, where the wall effects are appreciable, we add the contribution of interaction of surface molecules with the walls. If we consider a fluid in a confined-nano system the surface free energy contribution to the Helmholtz free energy may be represented by $U\psi N_{\psi}$, where $U$ is the wall-molecule average interaction energy per unit molecule per unit surface, $N_{\psi}$ is the number of molecules on the surface and $\psi$ is the surface area. Then, the Helmholtz free energy of the nano confined fluid systems is

$$A = N\mu - \sum_i p_{ii} d\delta_{ii} V + U\psi N_{\psi}$$ (9)

By differentiation of this equation we get,

$$dA = -SdT - \sum_i p_{ii} d\delta_{ii} V + \mu dN + UN_{\psi} d\psi .$$ (10)

Now, depending on the geometry of the confinement of the nano system under consideration the solution of this equation for pressure will vary. In what follows application of this equation for the two cases of cylindrical (nanotube) and slit-like (nanoslit) nano systems are reported.

## Phase Transition in A Nanotube:

To obtain the pressure coordinates for a confined fluid in a cylindrical nanopore, we take partial derivative of the Helmholtz free energy with respect to volume at constant $T$, $N$ and $L_z$. Therefore by inserting $\psi = 2\pi r L_z$, $d\psi = 2\pi (r dL_z + L_z dr)$ into the Helmholtz free energy, Eq.s (9) and (10), and using,

$$-p_{ii} = \frac{1}{V}\left(\frac{\partial A}{\partial \varepsilon_{ii}}\right)_{T,N,A},$$
(11)

we can get the pressure coordinates for the confined fluid in a nanotube. For the $z$-coordinate of the pressure vector we get

$$-p_{zz} = \left(\frac{\partial A}{\partial V}\right)_{T,N,r} = \frac{-NkT}{V-Nb} - \frac{kTN^2 b}{(V-Nb)^2} + \frac{2aN^2}{V^2} - p_{zz} - V\left(\frac{\partial p_{zz}}{\partial V}\right)_{T,N}, r + \frac{2U\,N_\psi}{r}$$
(12)

To get the $r$-coordinate of the pressure tensor (assuming angle-independent pressure), since $V = \pi r^2 L_z$ and $r = \sqrt{\frac{V}{\pi L_z}}$ we conclude $\psi = 2(\pi L_z V)^{1/2}$. Then by replacing this expression in Eq. (9) and differentiation of the Helmholtz free energy with respect to volume at constant $T$, $N$ and $L_z$ we get,

$$-p_{rr} = (\frac{\partial A}{\partial V})_{T,N,Lz} = \frac{-NkT}{V-Nb} - \frac{kTN^2b}{(V-Nb)^2} + \frac{2aN^2}{V^2} - p_{rr} - V(\frac{\partial p_{rr}}{\partial V})_{T,N,Lz} + \frac{(\pi L_z)^{1/2}UN_\psi}{V^{1/2}}, \quad (13)$$

of course, by assuming the radial symmetry of the cylindrical nanopore. Now let us write the variables appearing in the above two equations in reduced form. Then, by rearranging them we get the following formulas:

$$(\frac{\partial P_{zz}^*}{\partial v^*})_{T,N,r} = \frac{-T^*}{v^*(v^*-1)} - \frac{T^*}{v^*(v^*-1)^2} + \frac{2}{v^{*3}} + \frac{2U^*N_\psi}{r^*v^*}, \quad (14)$$

$$(\frac{\partial P_{rr}^*}{\partial v^*})_{T,N,Lz} = \frac{-T^*}{v^*(v^*-1)} - \frac{T^*}{v^*(v^*-1)^2} + \frac{2}{v^{*3}} + \frac{(\pi L_z^*)^{1/2}U^*N_\psi}{N^{1/2}v^{*3/2}}, \quad (15)$$

where,

$$p^* = \frac{pa}{b^2}, T^* = \frac{kTb}{a}, v^* = \frac{V}{Nb}, U^* = \frac{Ub^{5/3}}{a}, r^* = \frac{r}{b^{1/3}}.$$ and $k$ is the Boltzmann constant. Now, by integrating Eq.s (14) and (15) with the boundary condition that the wall effects will diminish when $N \rightarrow \infty$ or $r \rightarrow \infty$, $p_{rr}^* = p_{zz}^*$, we get:

$$P_{zz}^* = \frac{T^*}{v^*-1} - \frac{1}{v^{*2}} + \frac{2U^*N_\psi}{r^*}\ln v^*, \quad (16)$$

$$P_{rr}^* = \frac{T^*}{v^*-1} - \frac{1}{v^{*2}} - \frac{2U^*N_\psi}{N^{1/2}r^*}. \quad (17)$$

By using Eqs. (16) and (17) we have plotted $p_{zz}^*$ and $p_{rr}^*$ versus $v^*$ and we have compared them with macroscopic pressure, $p_{Macro}^*$, for $T^* = 0.25$, a subcritical isotherm, and for $\frac{U^*N_\psi}{r^*} = -0.004$ and $-0.2$ (two different assumed values) in Figures (2) and (3), respectively. All these numerical values chosen are arbitrary, but indicate two distinctly different contributions due to wall effects. As it is clear from Figures (2) and (3), both of the pressures can predict the vdW phase transition, s-shaped loops, and therefore, using the Maxwell construction principle (equality of temperatures, pressures and chemical potentials in the two phases), the properties of phases in equilibrium may be obtained. Note that in this analysis we have assume the pressure for a specific confined fluid depends on three variables ($T^*, v^*, y = \frac{U^* N_\psi}{r^*}$). Of course it is obvious that $y$ is the third degree of freedom. In other words, if $T^*$, $v^*$ and $y$ are defined for a confined fluid then the state of the fluid is completely defined.

As it is clear from Figures (2) and (3), $p_{zz}$ and $p_{rr}$ of nano confined van der Waals fluid are different from one another and from the macroscopic pressure. The difference between $p_{zz}$ and $p_{rr}$ is the result of the different local densities in different parts of the cylindrical nanopore. By comparing Eq.s (14) and (15) it is obvious that $(\frac{\partial P_{zz}^*}{\partial v^*})_{T,N,r}$ is different from $(\frac{\partial P_{rr}^*}{\partial v^*})_{T,N,Lz}$. Therefore, phase transition (condensation or evaporation in the present case) of a confined fluid may be different in different directions. Accordingly, we may also conclude the density is not uniform in confined fluids as it has been confirmed already in the literature[2,6,14,23]. By considering the local density we are able to interpret the existence of phase-transition fragmentation[14,24] in nano confined fluids.

Figure (4) shows the coexistence curve for $p_{zz}$ and compared it with the coexistence curve which has been obtained for a macroscopic system using the vdW eos. As it is also clear from Figure (4), for a fluid inside a nanopore shape of the coexistence curve is sensitive to wall effects (variable $y$). This is consistent with the simulation results as it has been reported in the literature[25].

We have also calculated the critical point behavior for confined vdW fluid in a nanotube and compared it with the literature molecular dynamics (MD) simulation data for confined water in nanotube as reported in Figure (5). To obtain the critical point properties, we derived the first- and second-partial derivative of $p_{zz}$ with respect to volume at constant $T$ and $N$ and set them equal to zero as follows,

$$(\frac{\partial p_{zz}^*}{\partial v^*})_{T,N} = \frac{-T^*}{(v^*-1)^2} + \frac{2}{v^{*2}} + \frac{2U^* N_\psi}{r^* v^*} = 0 \tag{18}$$

$$(\frac{\partial^2 p_{zz}^*}{\partial v^{*2}})_{T,N} = \frac{2T^*}{(v^*-1)^3} - \frac{6}{v^{*4}} - \frac{2U^* N_\psi}{r^* v^{*2}} = 0 \tag{19}$$

By solution of the above two equations we derive the following expressions for the dimensionless critical temperature, volume and pressure

$$T_c^* = \frac{8}{27} - \frac{14}{9}x + \frac{38}{27}x(\frac{d}{3x} + \frac{x+6}{3d} - \frac{1}{3}) - \frac{1}{9}x^2(\frac{d}{3x} + \frac{x+6}{3d} - \frac{1}{3})$$
$$- (\frac{4}{27}x + \frac{1}{9}x^2)(\frac{d}{3x} + \frac{x+6}{3d} - \frac{1}{3})^2, \tag{20}$$

$$v_c^* = \frac{d}{3x} + \frac{x+6}{3d} - \frac{1}{3}, \tag{21}$$

and

$$p_{czz}^* = \frac{\frac{8}{27} - \frac{14}{9}x + \frac{38}{27}x(\frac{d}{3x} + \frac{x+6}{3d} - \frac{1}{3}) - (\frac{4}{27}x + \frac{1}{9}x^2)(\frac{d}{3x} + \frac{x+6}{3d} - \frac{1}{3})}{\frac{d}{3x} + \frac{x+6}{3d} - \frac{4}{3}}$$

$$- \frac{1}{(\frac{d}{3x} + \frac{x+6}{3d} - \frac{1}{3})^2} + \frac{2U^* N_\psi \ln(\frac{d}{3x} + \frac{x+6}{d} - \frac{1}{3})}{r^*}, \tag{22}$$

where

$$x \equiv \frac{2U^* N_\psi}{r^*}, \tag{23}$$

and

$$d \equiv [(-90 - x + 3\sqrt{6}\sqrt{\frac{3x^2 + 148x - 4}{x}})x^2]^{1/3}. \tag{24}$$

We have calculated and plotted the reduced critical temperature for confined fluid in nanotubes versus nanotube radius as it is reported in Figure (5). According to this figure the confined fluid reduced critical temperature increases with the nanopore radius ($r$) and it reaches to its macroscopic (bulk) value (at $r=\infty$). We have also compared our reduced critical temperature results for $y=-0.001$ with the literature values which are obtained from the MD simulation for confined water in cylindrical nanopores[26]. The closeness of our dimensionless results with the dimensionless simulation data of water are surprisingly good. In our calculations we have practically assumed the simple van der Waals potential model, Eq. (2), while for simulating water the sophisticated *TIP4P* potential model was used. The critical densities for $p_{zz}$ at different nanotube diameter ($r$) sizes are also reported in Figure (6). Comparison of Figures (5) and (6) indicates that the critical density reaches to the bulk critical density much faster than the critical temperature dose as the radius of the nanotube increases.

## Phase Transition in A Nanoslit:

The same procedure used to derive the equations of state for a fluid confined in a cylindrical nanopore is also used for a nanoslit. We define a nanoslit consisting of two parallel walls, infinite in the *x-y* plane separated by a nanoscale width $H$ in the $z$ direction. Therefore the value of $\psi$ in Eq. (9) is $\psi=l_x l_y$, and the Helmholtz free energy for fluid confined in a nanoslit is[1]:

$$A = Nf(T) - NkT\ln(V - Nb) + \frac{kTN^2 b}{V - Nb} - \frac{2aN^2}{V} - \sum_i p_{ii}V_{ii} + 2U^* l_x l_y \qquad (25)$$

Now analytic expressions for pressure coordinates, $p_{xx}$, $p_{yy}$ and $p_{zz}$, can be obtained using Eq. (11), the partial derivative of $A$ with respect to $V_{xx}, V_{yy}$ and $V_{zz}$, respectively. The results are:

$$p_{xx}^* = p_{yy}^* = \frac{T^*}{v^* - 1} - \frac{1}{v^{*2}} + \frac{2U^* N_\psi}{H^*} \ln(v^*) \qquad (26)$$

$$p_{zz}^* = \frac{T^*}{v^* - 1} - \frac{1}{v^{*2}} \qquad (27)$$

As it is shown, the expression for $p_{zz}^*$ is equivalent to the macroscopic expression for pressure and the expressions for $p_{xx}^*$ and $p_{yy}^*$ are identical to one another.

Figure (7) Shows the plot of $p_{xx}^*$ versus reduced molar volume for $T^* = 0.25$ and $y = -0.001$. We have also plotted the coexistence curve for $p_{xx}^*$ by using the Maxwell construction principle and we have compared the results with the macroscopic vdW fluid coexistence curve in Figure (8) for $y = -0.001$. According to this figure the confined fluid coexistence curve is sensitive to the wall effects in nanolits, as expected. We have also calculated $T_c^*$ and $v_c^*$ for $p_{xx}^*$ at various nanoslit widths and the results are compared with the dimensionless literature MD simulation data of confined water in nanoslit as reported in Figure (9). According to Figure 9 the dimensionless results of predictions for $y = -0.001$ are quite close to the dimensionless literature MD simulation data of water confined in a nanoslit. We have also calculated the critical density for $p_{xx}^*$ versus nanoslit width and the results are reported in Figure (10). The prediction in this figure is quite similar to the prediction for fluid in a nanotube as shown in Figure (6).

## Conclusions and Discussion:

In this report we have introduced an analytic model for prediction of phase transitions of fluids confined in nano systems. Specifically the model is applied to the cases of fluid confined in nanotube and nanoslit. The proposed model is based on the van der Waals equation of state. It is proven that the van der Waals equation of state is valid for small as well as large systems, which makes it applicable for any size fluid system.

It is shown that small / nano systems which are not in the thermodynamic limit have, at least, an additional degree of freedom unknown to large systems in thermodynamic limit. This is in agreement with the initial findings of T.L. Hill[6]. Our results are also indicative of the fact that the density is not uniform in nanopores. Variation of local density for a fluid in a nanopore or nanoslit may be a cause for its fragmentation[24] during phase transition in small systems. We have demonstrated that critical temperature and critical density of a confined fluid vary with the size of its confinement and they increase as the size of the confinement increases. The similarities between the dimensionless results of the simple vdW eos for a confined fluid and dimensionless result of the MD simulation for confined water[27] are surprisingly good.  It is also demonstrated that the coexistence curve for nano confined fluids is sensitive to the nature of confinement surface in agreement with the literature data[27].

# Nomenclature

| | |
|---|---|
| $A$ | Helmholtz free energy |
| $a$ | van der Waals equation constant |
| $b$ | van der Waals equation constant |
| $d$ | differential |
| $E$ | internal energy |
| $f$ | function |
| $H$ | nanoslit width |
| $k$ | Boltzmann constant |
| ln | natural log |
| $L$ | length |
| $n$ | intermolecular potential exponent |
| $N$ | number of molecules |
| $p$ | pressure tensor |
| $p$ | pressure tensor component |
| $P$ | total pressure |
| $r$ | intermolecular distance |
| $S$ | total entropy |
| $T$ | absolute temperature |
| $U$ | wall-molecule average interaction energy per unit molecule per unit surface |
| $v$ | molecular volume |
| $V$ | total volume |
| $y$ | $= \dfrac{U^* N_\psi}{r^*}$ |

## Greek Letters

| | |
|---|---|
| $\partial$ | partial derivative |
| $\delta$ | deformation |
| $\varepsilon$ | intermolecular energy parameter |
| $\sigma$ | intermolecular length prameter |
| $\mu$ | molecular chemical potential |
| $\pi$ | =3.1415927 |
| $\phi$ | intermolecular pair potential function |
| $\psi$ | surface area |
| $\Sigma$ | summation |

## Superscripts

| | |
|---|---|
| $ig$ | ideal gas |
| $*$ | dimensionless |

## Subscripts

| | |
|---|---|
| $i, j$ | mlolecule numbers |
| $r$ | radial coordinate |
| $x,y,z$ | Cartesian coordinates |
| $\psi$ | on the surface |

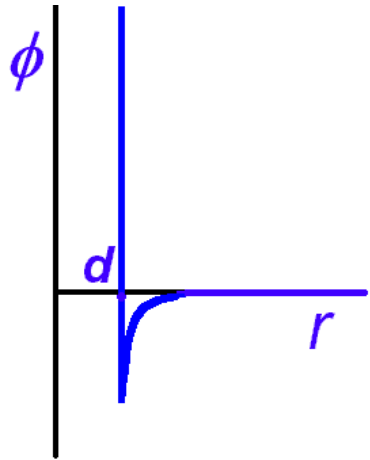

*Figure (1) – Intermolecular potential function – the basis of the van der Waals equation of state.*

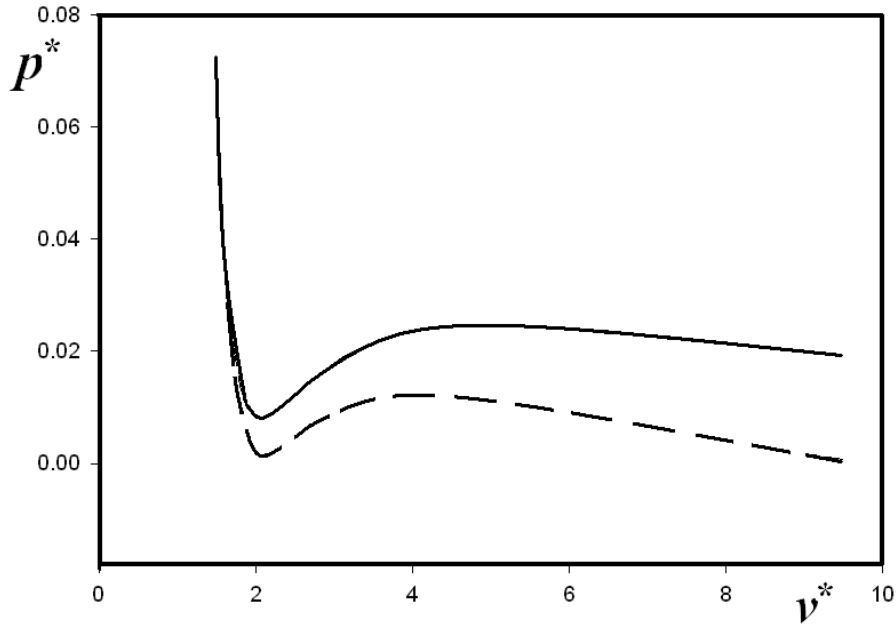

*Figure (2) - The reduced pressures, $p^*_{zz}$ (− − −), $p^*_{rr}$ (——) and $p^*_{Macro}$ (——), versus the reduced volume $v^*$ for a fluid inside a nanotube. In this figure $T^*=0.25$ and $y=\dfrac{U^*N_\psi}{r^*}=-0.004$. Note that in the case of this figure values of $p^*_{rr}$ and $p^*_{Macro}$ are graphically indistinguishable from one another.*

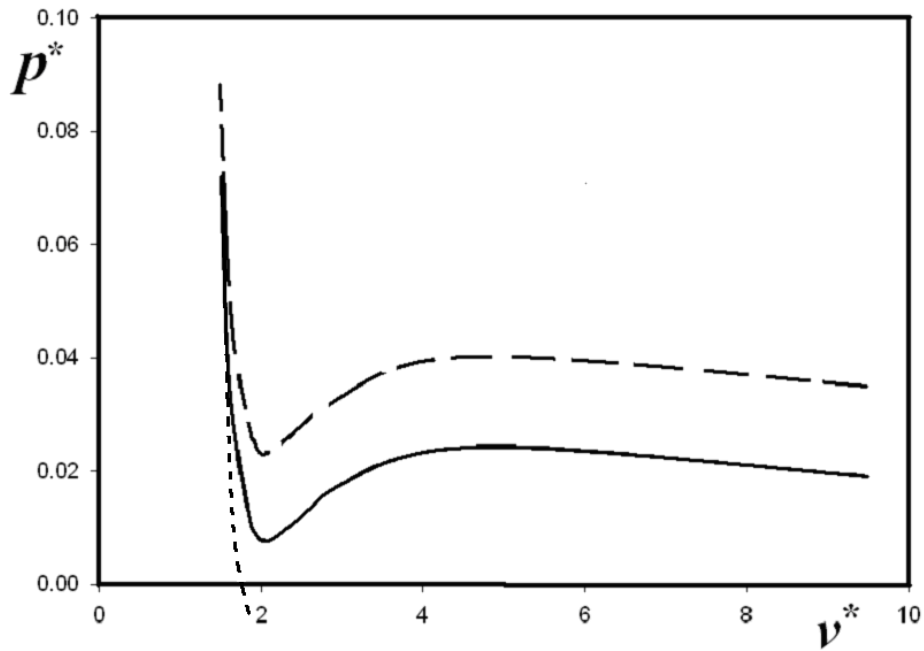

*Figure (3) - The reduced pressures $p^*_{zz}$ (– – –), $p^*_{rr}$ (-------) and $p^*_{Macro}$ (——), versus the reduced volume $v^*$ for a fluid inside a nanotube. In this figure $T^*=0.25$ and $y = \dfrac{U^* N_\psi}{r^*} = -0.2$.*

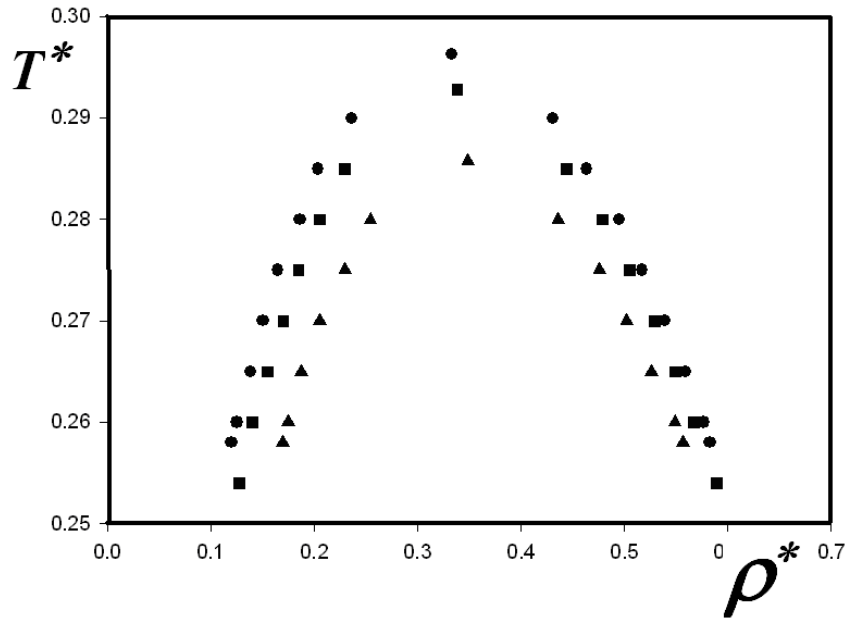

*Figure (4) - The coexistence curves (dimensionless temperature versus dimensionless density) for the confined vdW fluid inside a nanotube as obtained from $p^*_{zz}$, Eq (12), and compared with the coexistence curve for the vdW fluid in macroscopic scale.  In this figure the data shown by squares are for $y = -0.001$, the triangles are for $y = -0.004$ and circles are for the macroscopic system.*

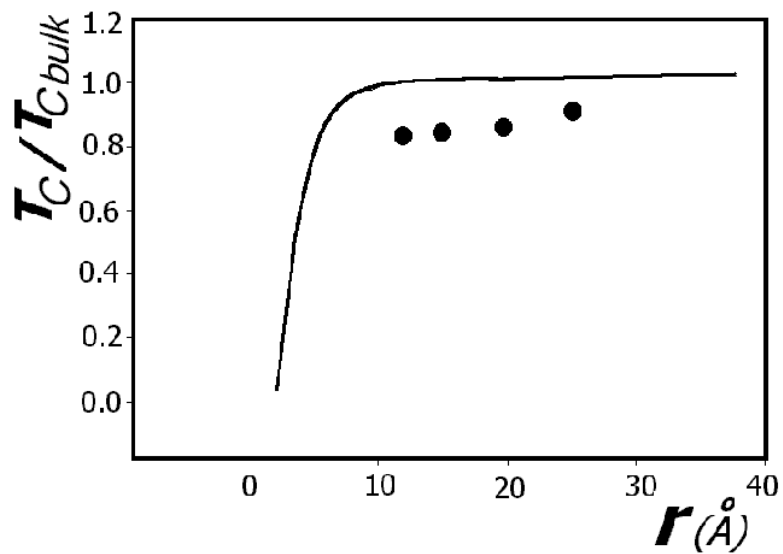

*Figure (5) -The ratio of critical temperature of the vdW fluid in nanotube over the macroscopic (bulk) critical temperature ($T_{Cz}$ /$T_{Cbulk}$) as a function of nanotube radius (r). The solid line is calculated based on Equation (12) and for $y = -0.001$. The solid circles are the MD simulation data for confined water [25] in nanotube.*

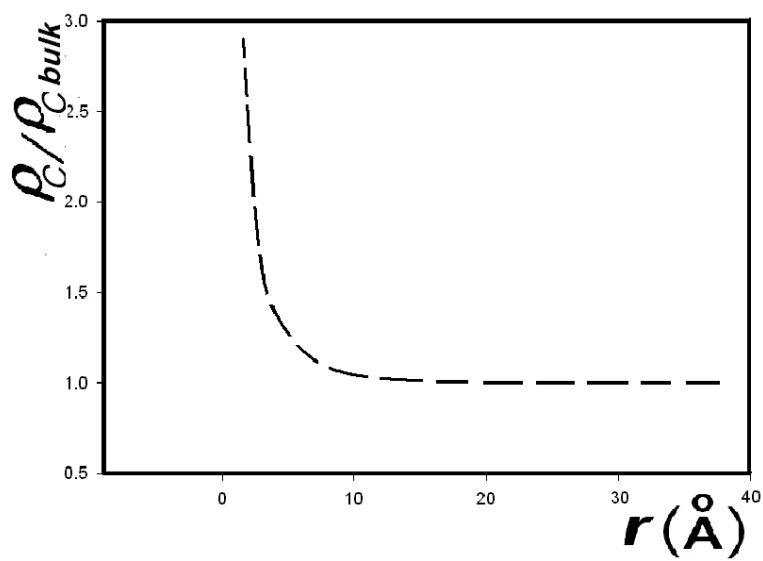

*Figure (6) - The ratio of critical density of fluid in nanotube over the macroscopic (bulk) critical density ( $\rho_{Czz} / \rho_{Cbulk}$ ) as a function of nanotube radius (r) calculated based on Equation (12) and for y = −0.001.*

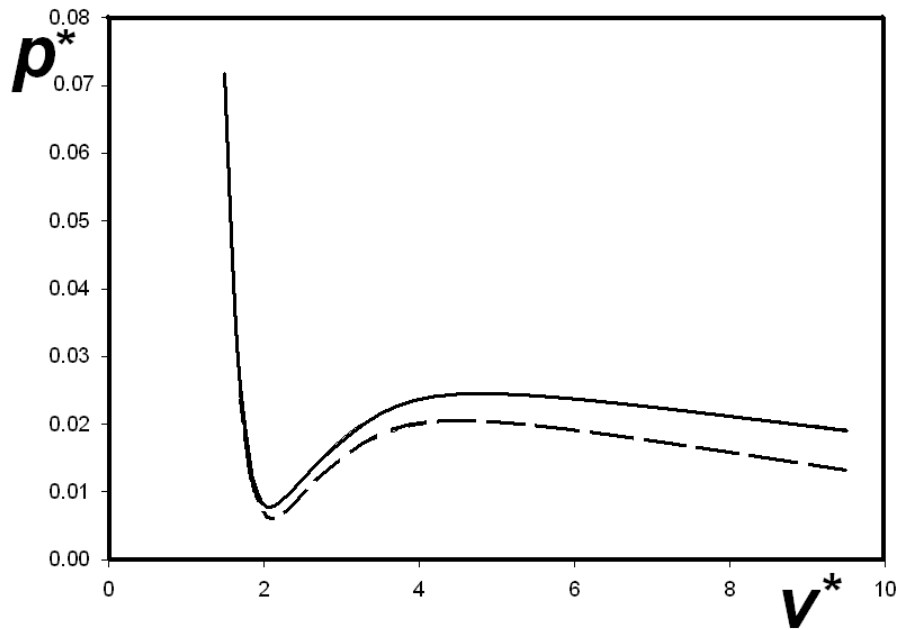

Figure (7) - The reduced pressures $p_{xx}^* = p_{yy}^*$ $(---)$, $p_{zz}^*$ $(\textemdash)$ and $p_{macro}^*$ $(\textemdash)$ versus the reduced volume $v^*$ for the confined vdW fluid inside a nano-slit. In this figure $T^* = 0.25$ and $y = -0.001$. Note that in the case of this figure values of $p_{zz}^*$ and $p_{Macro}^*$ are graphically indistinguishable from one another.

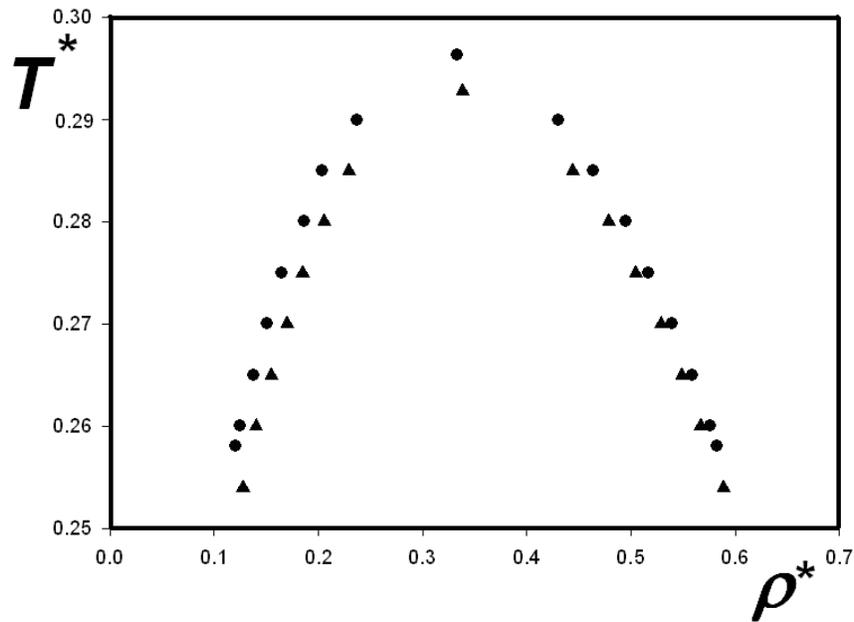

*Figure (8) - The coexistence curves (dimensionless temperature versus dimensionless density) for the confined vdW fluid inside a nanoslit as obtained from $p^*_{xx} = p^*_{yy}$, Eq (26), and compared with the coexistence curve for the vdW fluid in macroscopic scale. In this figure the data shown by triangles are for $y = -0.001$ and the data shown by circles are for the macroscopic system.*

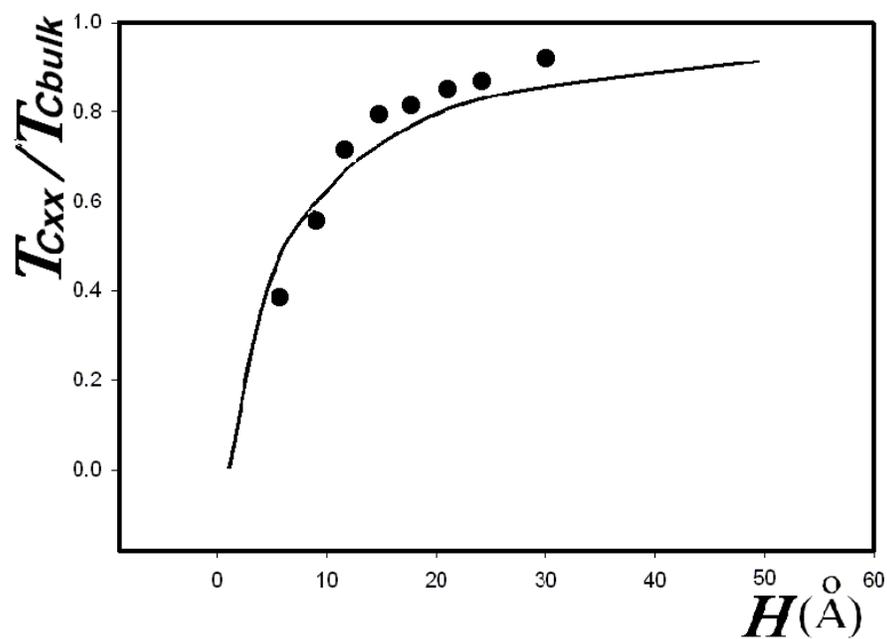

*Figure (9) - The ratio of critical temperature of a fluid in nanoslit over the macroscopic (bulk) critical temperature ($T_{Czz}/T_{Cbulk}$) for different sizes of nano-slit width (H). The solid line is results for the vdW fluid calculated based on Equation (26) and for y = −0.001. The solid circles are the MD simulation results [25] for confined water in nanoslit.*

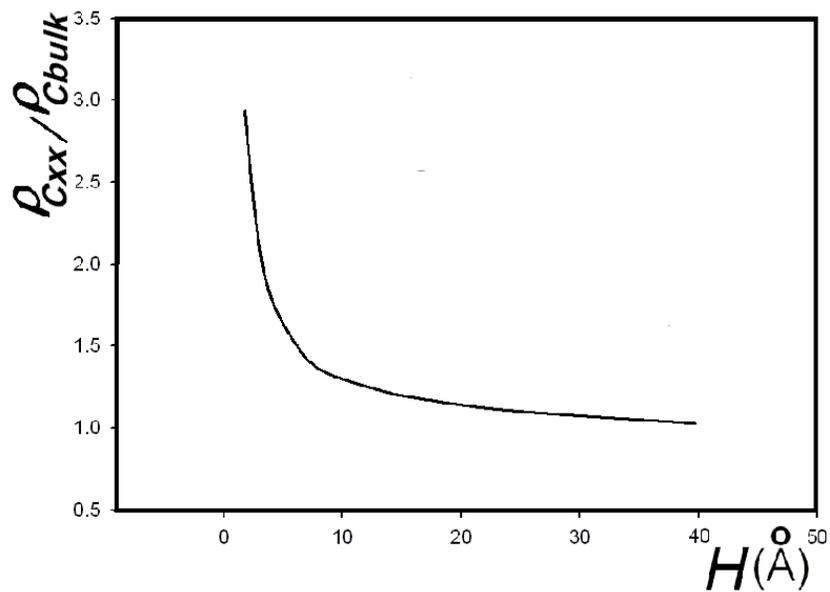

*Figure (10) - The ratio of critical density of the vdW fluid in nanoslit over the macroscopic (bulk) vdW fluid critical density ( $\rho_{Cxx} / \rho_{Cbulk}$ ) for different sizes of nanoslit width (H) calculated based on Equation (26) and for $y = -0.001$.*